# Andreev reflection spectroscopy of the heavy-fermion superconductor $CeCoIn_5$ along three different crystallographic orientations

Wan Kyu Park,[a,*] Laura H. Greene,[a] John L. Sarrao,[b] Joe D. Thompson[b]

[a]*Department of Physics and Frederick Seitz Materials Research Laboratory, University of Illinois at Urbana-Champaign, Urbana, IL 61801, USA*

[b]*Los Alamos National Laboratory, Los Alamos, NM 87545, USA*

**Abstract**

Andreev reflection spectroscopy has been performed on the heavy-fermion superconductor (HFS) $CeCoIn_5$ single crystals along three different crystallographic orientations, (001), (110), and (100), using Au tips as counter-electrodes. Dynamic conductance spectra are reproducible over wide temperature ranges and consistent with each other, ensuring the spectroscopic nature. Features common to all directions are: i) asymmetric behaviors of the background conductance, which we attribute to the emerging coherent heavy-fermion liquid; ii) energy scales (~1 meV) for conductance enhancement due to Andreev reflection; iii) magnitudes of enhanced zero-bias conductance (10 - 13 %). These values are an order of magnitude smaller than the predicted value by the Blonder-Tinkham-Klapwijk (BTK) theory, but comparable to those for other HFSs. Using the *d*-wave BTK model, we obtain an energy gap of ~ 460 μeV. However, it is found that extended BTK models considering the mismatch in Fermi surface parameters do not account for our data completely, which we attribute to the shift of spectral weight to low energy as well as to the suppressed Andreev reflection. A qualitative comparison of the conductance spectra with calculated curves shows a consistency with $d_{x2-y2}$-symmetry, providing the first spectroscopic evidence for the order parameter symmetry and resolving the controversy over the location of the line nodes.

## 1. Introduction

The heavy-fermion superconductor (HFS) family $CeMIn_5$ (M = Co, Rh, Ir) has drawn much attention since they exhibit a variety of interesting physical phenomena ([1] & Refs. therein). As for the superconducting order parameter of $CeCoIn_5$, many transport and thermodynamic measurements have indicated the existence of line nodes [1], implying an unconventional pairing symmetry. In particular, The fourfold symmetry in the magnetic field-angle dependent thermal conductivity [2] and specific heat [3] measurements in the *ab*-plane have shown a good agreement with *d*-wave pairing state.

However, it has remained controversial whether the line nodes are located along (100) or (110) direction. Recently, Vorontsov and Vekhter [4] have reported that theoretical calculations taking into account the competition between transport scattering rate and density of states could resolve this controversy. Their results are in favor of the $d_{x2-y2}$ symmetry in $CeCoIn_5$, hence, line nodes along the (110) direction.

Although field-angle dependent experiments are useful tools for probing the anisotropy of the order parameter amplitude, they are not sensitive directly to the phase change of the order parameter. Meanwhile, phase-sensitive measurements, including Josephson interferometry [5] and scanning SQUID microscopy [6], have proved to be decisive in detecting such phase change directly. It has also been well established that single particle tunneling spectroscopy can be a phase-sensitive technique via detection of zero-energy Andreev bound state (ABS) on a

* Corresponding author. Tel.: +1-217-265-5010; fax: +1-217-244-8544; e-mail: wkpark@uiuc.edu.

*d*-wave superconductor surface whose normal is along the nodal direction [7,8].

Owing to its technical simplicity, point-contact spectroscopy (PCS) has been frequently adopted as an alternative tool to conventional tunneling spectroscopy. It is true that PCS has been playing an important role in determining energy gaps of novel superconductors [9]. However, detecting the order parameter phase via PCS is rather difficult since generally it measures charge transport in the intermediate regime between Andreev reflection (AR) [10] and tunneling, thus making data analysis complicated. Recent PCS results [11-14] on $CeCoIn_5$ and debates [15-17] over how to interpret them have clearly shown such difficulty. The conductance spectra obtained from PCS are very sensitive not only to surface states in the tip and sample but also to geometric factors including contact size and pressure. To ensure that they are representative of the spectroscopic information intrinsic to the material under study, PCS performed in different crystallographic orientations [13,14] provides a crucial diagnostic. Here we report such measurements along three different directions, supporting the $d_{x2-y2}$ symmetry in $CeCoIn_5$.

## 2. Experiments

The PCS is carried out using our home-built rig [18]. A point contact junction (PCJ) is made by bringing an electro-chemically prepared Au tip into contact with a single crystal $CeCoIn_5$ via a combined adjustment of a fine screw and piezoelectric arms. Detailed experimental procedures can be found elsewhere [14,18]. Here we focus on preparation and characterization of samples with different crystallographic orientations.

For (001) PCJs, as-grown single crystals with shiny and smooth surfaces are chosen and used after chemical etching in HCl and cleaning. For (110) & (100) PCJs, a thick rectangular-shaped single crystal is embedded into epoxy, with the perpendicular direction adjusted along (110) or (100) axis. After the epoxy is hardened, it is cut in the lateral direction so that the exposed crystal surface is oriented along each direction. Then, it is polished using alumina/diamond lapping films and silica colloidal suspensions with the particle size down to 25 nm. The polished surface appears very smooth and mirror-like shiny when examined using a high-resolution optical microscope.

The orientation of each sample is checked by x-ray diffraction as shown in Fig. 1. The peaks marked by * are identified as due to the epoxy. Figure 1(a) is for a c-axis oriented crystal, which is not embedded into epoxy. It is clearly seen that sample surfaces in Fig. 1(b) and (c) are oriented along (110) and (100) direction, respectively, as designed. No peaks other than crystals and epoxy peaks are detected, implying that prepared samples are made of single phase of $CeCoIn_5$ with well-defined crystallographic axes. Nonetheless, to remove any possibly remaining indium precipitate, these sample surfaces are etched by HCl and cleaned.

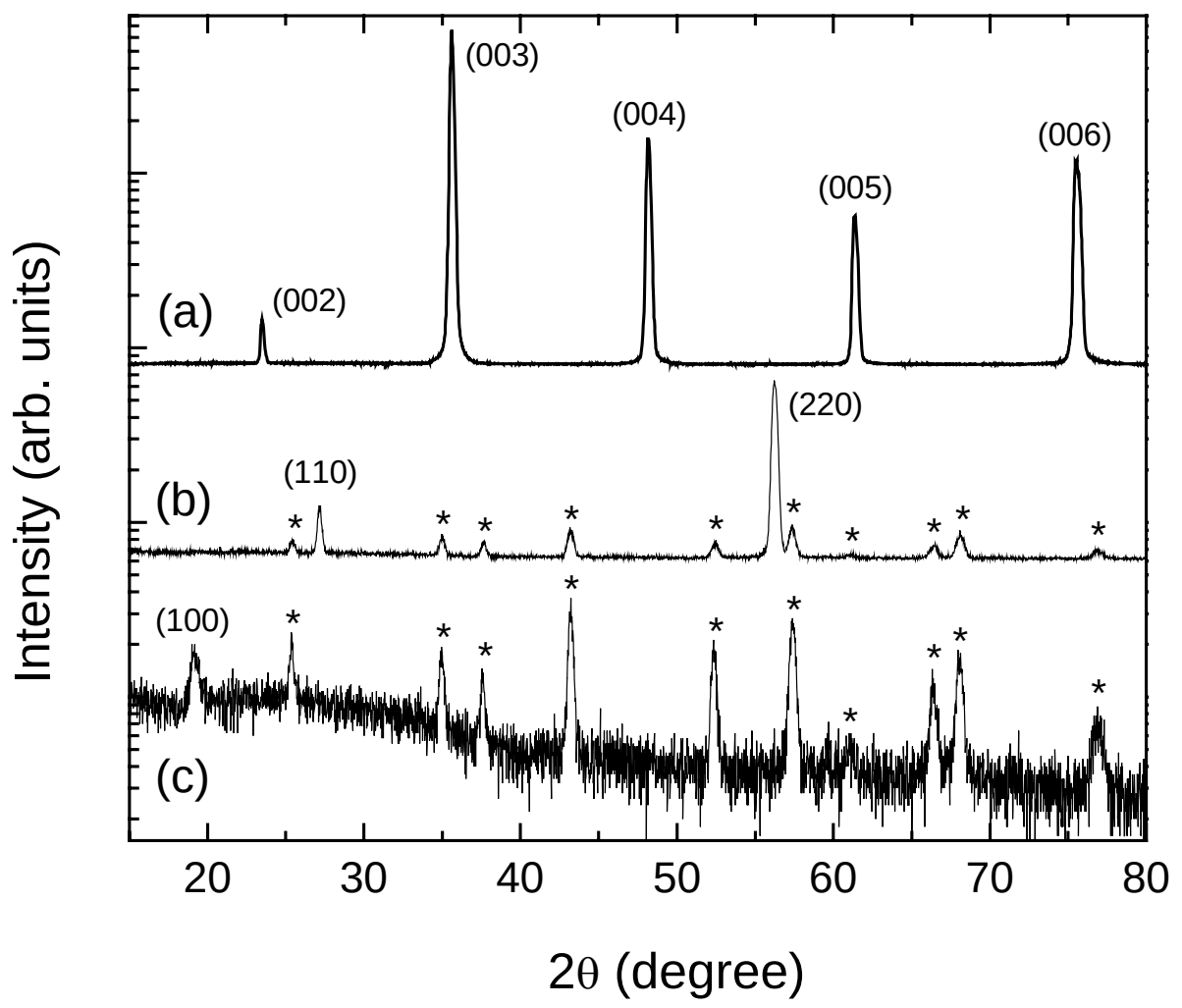

Fig. 1. X-ray diffraction data for $CeCoIn_5$ single crystals. (a) Only (00*l*) peaks are detected, indicating the *c*-axis orientation. (b) & (c): crystals embedded into epoxy, cut, and polished. The peaks marked with * are due to epoxy. (b) (110) orientation. (c) (100) orientation.

Dynamic conductance spectra are taken using the standard four probe lock-in technique as a function of bias voltage, temperature down to 400 mK, and magnetic field up to 9 Tesla.

## 3. Results and discussion

Here we present conductance spectra for PCJs along three directions at two temperatures only, the lowest and a higher. Full sets of temperature dependent data for (001) and (110) PCJs have been reported elsewhere [13,14].

### *3.1. Spectroscopic nature*

As mentioned above, it is essential to establish whether measured data are spectroscopic or not before attempting to perform extensive analysis and to interpret it as indicative of intrinsic properties of the material in question. The spectroscopic nature of a PCS result has been claimed frequently from the contact being in the Sharvin limit [19]. However, this is just good for giving a rough idea since one does not directly measure the contact size and materials properties in the contact area can be different from those in bulk.

Here we claim that our conductance spectra are representing the intrinsic electronic properties of $CeCoIn_5$. Contact sizes estimated from measured contact resistances (1 – 5 Ω) at the highest bias voltages fall to the Sharvin limit. More importantly, the PCJs along three different crystallographic directions exhibit consistent spectra with each other and theoretical predictions, as discussed below.

### *3.2. Consistency in the conductance spectra*

Dynamic conductance spectra, normalized at – 2 mV, are displayed in Fig. 2. It was reported previously that an asymmetry in the background conductance starts developing around the heavy-fermion coherence temperature, $T^*$ ~ 45 K, remaining almost constant below $T_c$ (2.3 K) [13,14]. We attribute this asymmetry to the emerging coherent heavy-fermion liquid as in the two-fluid model proposed by Nakatsuji, Pines, and Fisk [20].

Below $T_c$, conductance enhancement due to AR is observed in the sub-gap region (~±1 mV), as seen in Fig. 2. It is quantified by further normalization of the data with the constant background conductance [13,14]. The amount of conductance enhancement at zero-bias is ~13% and ~12%, for (001) and (110) PCJ, respectively. A rough estimate gives about 10% for (100) PCJ from Fig. 2(c). Thus, the zero-bias conductance enhancement is ~10-13 % in all three directions, an order of magnitude smaller than the prediction (100%) of the Blonder-Tinkham-Klapwijk (BTK) theory [21] and smaller than in conventional superconductors by several factors of magnitude [9]. This reduced AR signal has also been observed from the PCS on other HFS ([22] & Refs. therein). We believe there must be a common mechanism for these observations.

We have carried out an extensive analysis to quantify the AR conductance using extended BTK models [13,14]. Although the *d*-wave BTK model [23] could account for our data to some extent, giving rise to an estimated energy gap of ~ 460 μeV at 400 mK, it cannot fit to the whole set of temperature dependent conductance spectra. We also attempted to fit the data by taking into account effects of the mismatch in Fermi surface parameters. However, such calculations just give rise to the usual BTK conductance with proper scaling of the parameters [24,25]. All these failures might be due to the spectral weight shift to lower energy as well as to the suppressed AR at the HFS interface [13,14].

### *3.3. Evidence for $d_{x2-y2}$ wave symmetry*

Conductance curves at the lowest temperatures appear flat around zero-bias for (001) and (100) PCJs, whereas exhibit contrasting features for (110) PCJ in Fig. 2(b): slope change and resulting cusp-like shape near zero-bias. This difference can be detected even at higher temperatures, albeit slightly: the (001) PCJ conductance curve still maintains small flat region and the (100) PCJ curve is rounded due to a thermal smearing effect. On the other hand, the (110) PCJ curve still shows a cusp-like or triangular shape at a comparable temperature. For the full sets of temperature-evolved data, see Refs. 13 & 14.

For a qualitative understanding of these contrasting behaviors, we calculate conductance curves for junctions with arbitrary effective barrier strength, $Z_{eff}$ [23]. Figure 3 shows zero-temperature conductance curves for a junction whose normal is along the anti-nodal direction (Fig. 3(a)) and the nodal direction (Fig. 3(b)). Since the shapes of measured conductance curves fall to the AR regime rather than the tunneling regime, we focus on the small-$Z_{eff}$ limit. The AR-like conductance curve for the (001) PCJ can be understood by considering a large tunneling cone for small $Z_{eff}$. The $Z_{eff}$ =0 curve is the same for both nodal and anti-nodal junctions, whereas the shape changes in different manners with increasing $Z_{eff}$. Provided that the quasiparticle lifetime broadening factor is negligibly small, the shape of an experimental conductance curve at a sufficiently low temperature is determined by $Z_{eff}$. Because of unmatched Fermi velocities, we can expect $Z_{eff}$ to be non-zero in most normal-metal/HFS contacts [13,14], even after considering the theory by Deutscher and Nozières [26]. Then, junctions along nodal and anti-nodal directions should show opposite behaviors near zero-bias, as shown in Fig. 3. Comparing curves in Figs. 2 & 3, we can say that the (110) PCJ data are consistent with a nodal junction, while the (100) PCJ data are consistent with an anti-nodal junction. The cusp-like conductance curve in Fig. 2(b) represents the observation of ABS. The reason why it does not appear as a sharp peak at zero-bias as in a tunnel junction [7,8] is because the zero-energy ABS is smeared to finite energy due to small $Z_{eff}$ [23].

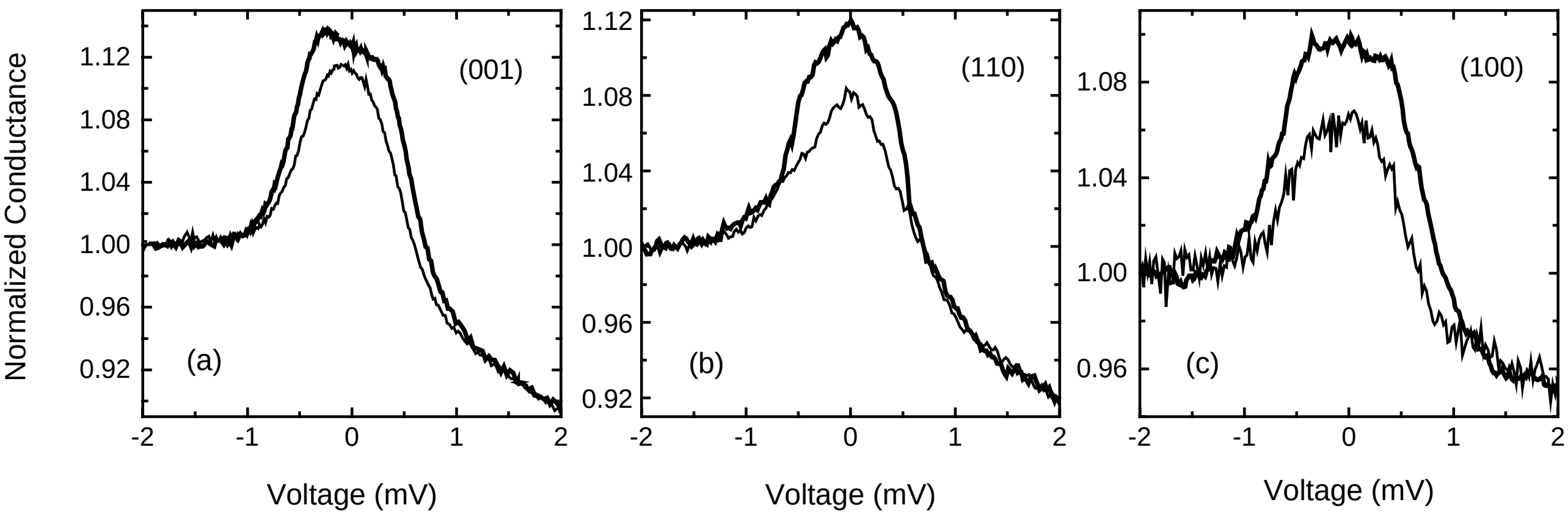

Fig. 2. Comparison of the dynamic conductance spectra of $CeCoIn_5$ along different crystallographic orientations: (a) (001), (b) (110), and (c) (100). Thick lines are taken at the lowest temperatures, 400 mK, 410 mK, and 420 mK, respectively. Thin lines are measured at higher temperatures, 1.52 K, 1.57 K, and 1.47 K, respectively. Note the cusp-like feature in the (110) data, in contrast to the other two curves for (001) and (100) junctions. This can be explained qualitatively using the *d*-wave BTK model, implying the $d_{x2-y2}$ wave pairing symmetry.

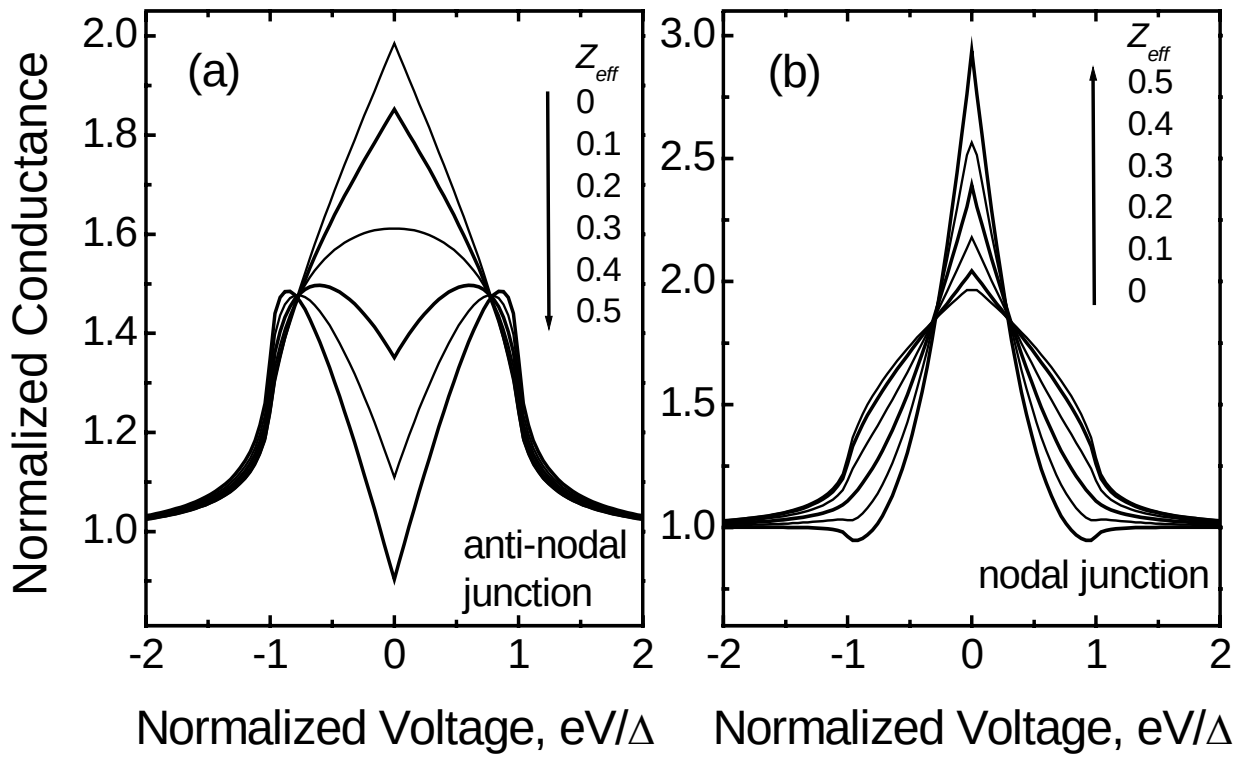

Fig. 3. Calculated conductance curves for normal-metal/*d*-wave superconductor junctions with arbitrary barrier strength, $Z_{eff}$. The junction normal is along the anti-nodal direction in (a), whereas along the nodal direction in (b). Note the contrasting behaviors between them with increasing $Z_{eff}$.

## 4. Conclusions and future directions

Dynamic conductance spectra along three different crystallographic orientations of $CeCoIn_5$ crystals have shown consistent behaviors in terms of the asymmetry in the background conductance, the amount of and the energy scale for the conductance enhancement at zero-bias. Quantitative analyses based on extended BTK models indicate that models considering only the mismatch in Fermi surface parameters cannot account for the full sets of conductance spectra. A model taking into account reduced AR, spectral weight shift, and the two-fluid behavior might provide better understanding of the Andreev conversion process at the normal-metal/HFS interface. Qualitative comparisons to *d*-wave BTK calculations provide the first spectroscopic evidence for the $d_{x2-y2}$-wave pairing symmetry, thus resolving the controversy over the location of line nodes in $CeCoIn_5$. More complete and quantitative analyses of the whole conductance spectra would be possible after a reasonable theoretical model is set up considering both reduced AR and shift of spectral weight. Our future directions also include continued studies of un-doped and Cd-doped $CeCoIn_5$, where at the 10% doping level, our preliminary temperature-dependent PCS data show characteristics consistent with antiferromagnetic and subsequent superconducting transitions observed in this material [27].

## Acknowledgments

W.K.P. and L.H.G. are grateful to A. J. Leggett, D. Pines, V. Lukic, and J. Elenewski for fruitful discussions. W.K.P. is thankful to X. Lu for his experimental help. This work is supported by the U.S. Department of Energy, Award DEFG02-91ER45439, through the Frederick Seitz Materials Research Laboratory and the Center for Microanalysis of Materials at the University of Illinois at Urbana-Champaign.